\documentclass[aps,prc,showpacs,11pt]{revtex4}
\usepackage{graphicx,epsfig}

\newcommand{ \be }{\begin{equation}}
\newcommand{ \ee }{\end{equation}}
\newcommand{ \bea }{\begin{eqnarray}}
\newcommand{ \eea }{\end{eqnarray}}
\newcommand{ \la }{\langle}
\newcommand{ \ra }{\rangle}

\newcommand{ \dpti }{\delta p_{t,1}}
\newcommand{ \dptj }{\delta p_{t,2}}
\newcommand{ \bp }{{\bf p}}
\newcommand{ \mpt }{{\la p_t \ra}}

\usepackage{color}

\begin{document}

\title{
Transverse radial expansion in nuclear collisions
and two particle correlations 
}

\author{Sergei A.~Voloshin}

\affiliation{Department of Physics and Astronomy, 
Wayne State University, Detroit, Michigan 48201}

\date{October 28, 2005} 

\begin{abstract}
At the very first stage of an ultra-relativistic nucleus-nucleus collision
new particles are produced in individual nucleon-nucleon collisions. 
In the transverse plane, all particles 
from a single $NN$ collision are initially located at the same position. 
The subsequent thermalization and transverse radial expansion 
of the system create
strong position-momentum correlations and
lead to characteristic rapidity, transverse momentum, and
azimuthal correlations among the produced particles. 
\end{abstract}

\pacs{25.75.Ld} 

\maketitle

The physics of the high energy heavy ion collisions attracts strong
attention of the physics community 
as creation of a new type of matter, the Quark-Gluon Plasma,
is expected in such collisions. 
During the last few years of Au+Au collisions at the BNL Relativistic
Heavy Ion Collider (RHIC) many new phenomena has been observed,
such as strong elliptic flow~\cite{star-flow1} and  
suppression of the high transverse momentum two particle back-to-back 
correlations~\cite{star-highpt}.
These observations strongly indicate that a dense partonic matter 
has been created in such high energy nuclear collisions.
Parton re-interactions lead to pressure build-up and
the system undergoes longitudinal and transverse expansion,
the latter leading to an increase in
the particle final transverse momenta.
The thermalization time of the system is estimated 
to be smaller than 1~fm/c~\cite{hydroreview}.
Here, we mostly discuss the effect of the transverse
{\em radial} expansion, neglecting a possible azimuthal dependence
in the transverse expansion velocity field, which is mostly important 
for the anisotropic flow study. 
Usually the transverse expansion is studied via detailed
analysis of single particle transverse momentum spectra, 
most often using thermal
parameterization suggested in~\cite{thermal-ssh}.
Transverse expansion also causes the transverse momentum dependence 
of the HBT radii~\cite{hbtradii-pt}.
In this paper we note that the transverse radial expansion should also lead
to characteristic rapidity, transverse momentum, and azimuthal angle
(long range, as opposed to the HBT like scale) 
two-particle correlations.
Those correlations are of a totally new type (compared to what 
is usually discussed)
having the nature of the correlations among particles placed at 
the same spatial location in the rapidly expanding dense medium. 

The origin of these correlations can be described as follows: at the
first stage of a AA collision many individual 
nucleon-nucleon collision happen. 
The strings stretched and new particles are produced (partons 
are freed; the exact mechanism of the particle production is 
not important for the presentation of the main idea of this paper).
Due to a rapid thermalization, the newly produced particles become
``frozen in'' the medium, preserving all the correlations between
them, e.g. the correlations due to the conservation laws of electric 
and/or baryon charges, strangeness, etc. 
The relative particle position (in the transverse plane) 
could be only  modified by particle
diffusion during the system evolution before the final freeze-out 
(for a more detailed discussion of the role of the thermalization time and
diffusion see later in the paper).   
Then the transverse expansion of the system 
would create strong position-momentum correlations 
in the transverse plane: in general, farther from 
the center axis of the system a particle is produced initially,
on average the larger (transverse, in radial outward direction) 
push it receives from other particles during the system evolution.
It is common~\cite{hydroreview} that the radial expansion 
collective velocity is parameterized by 
monotonically increasing function of the radial distance, though some
deviations from that could be possible at very large distances due to
low density at the periphery of the system.
All particles produced from the same string (from 
the same $NN$ collision) have initially 
the {\em same spatial} position in the transverse plane.
Consequently, all particle from the same string gets {\em on average} the
same push and thus become correlated.
Correlated pairs of particles, being produced at the same place, for
example containing $s-$ and $\bar{s}-$quarks, would be boosted in the
same direction and become azimuthally correlated, etc..

Note that in rapidity space the correlations due to described mechanism
could extend up to the length of the entire string.
For simplicity, in the most of the discussion below 
we assume the radial expansion of the system 
to be boost invariant.    
The 'push' is in the transverse direction, and on average does not affects
the longitudinal momentum component.

Based on the above mentioned ideas it appears a simple picture of 
$AA$ collisions,  which could be considered in some sense as a next order 
approximation to a simple superposition of independent $NN$ collisions
often use as a base line in correlation studies of $AA$ collisions.
In this picture, the particles produced in each independent collision 
are boosted in the radial direction depending to the location of
the collision in the transverse plane 
and the transverse expansion velocity profile.
It corresponds to non-zero space-momentum correlations in 
the transverse plane, $\la x_i, \Delta p_{t,i}\ra \ne 0 $ (
see~\cite{wiggle} as an example of the role of space-momentum correlations
in the development of directed flow).
As discussed below this picture leads to many distinctive phenomena, most 
of which can be studied by means of two (and many-) particle
correlations.

{\em Two particle transverse momentum correlations.}
The single particle spectra are affected by radial flow in such a way
that the effective slope as well as the mean transverse momentum
are  mostly sensitive to the {\em average} expansion velocity squared
(the validity of this approximation is discussed in more detail below) and to
much lesser extend to the actual velocity profile (dependence of the
expansion velocity on the radial distance from the center axis 
of the system).    
The two-particle transverse momentum correlations~\cite{vkr}, 
$\la p_{t,1} p_{t,2} \ra - \la  p_{t,1}  \ra \la p_{t,2} \ra
\equiv \la \delta p_{t,1} \, \delta p_{t,2} \ra$, would be
sensitive mostly to the {\em variance}
 in collective transverse expansion velocity, and thus are more
sensitive to the actual velocity profile.
Simple estimates show that the corresponding contribution could 
be comparable in magnitude to
the primordial correlations existed in  $NN$ collisions. 
For a rough estimate one can use a relation
\be
\la p_t\ra_{AA} \approx \la p_t \ra_{NN} + \alpha \la v^2 \ra,
\ee
where the coefficient $\alpha$ is of the order of typical particle
mass, and $v$ being the expansion velocity.
Then the correlation between transverse momenta of two particles would be
\be
\la \delta p_{t,1} \delta p_{t,2} \ra_{AA}  \approx
D_{N_{coll}}
(\la \delta p_{t,1} \delta p_{t,2} \ra_{NN} + \alpha^2 \sigma^2_{v^2}).
\label{edpt}
\ee 
The factor $D$, 
\be
D_{N_{coll}} =
\frac{\la n(n-1) \ra_{NN}}
{(N_{coll}-1) \la n \ra_{NN}^2 + \la n(n-1) \ra_{NN}},
\label{ed}
\ee 
takes into account the dilution of the
correlations due to a mixture of particles from $N_{coll}$ uncorrelated 
$NN$ collisions, and
that in an individual $NN$ collision the mean number of particle pairs,
$\la n(n-1) \ra_{NN}$, on average is larger than 
the mean multiplicity squared, $ \la n \ra_{NN}^2$ 
(see also~\cite{nu-method,gavin03}).
For a linear velocity profile, $\sigma^2_{v^2}=\la v^2 \ra^2/3$. 
Taking into account that the increase in the average transverse 
momentum due to the radial expansion
is of the order of 20--30\%, one concludes that the two terms in
Eq.~\ref{edpt} could be of the same order of magnitude:
an increase in the relative correlations, 
$ \alpha^2 \sigma^2_{v^2} / \la p_t \ra^2$, 
could be of the order of 1--3\%, similar to the value of
$\la \delta p_{t,1} \delta p_{t,2} \ra_{NN} / \la p_t \ra^2 \approx
(0.12)^2$,
measured at ISR~\cite{isrpp}. 
Thus, the correlations due to transverse expansion could be the major part 
in the observed centrality dependence of the mean $p_t$ 
fluctuations/correlations observed at the SPS and 
RHIC (for recent results, see~\cite{mptreviewQM2004}).

\begin{widetext}
More accurate estimates can be obtained employing a thermal model presented
in~\cite{thermal-ssh}. 
In this approach particles are produced by
freeze-out of the thermalized matter at temperature $T$, approximated 
by a boosted Boltzmann distribution. 
Assuming boost-invariant longitudinal expansion and freeze-out 
at constant proper time, one finds, up to irrelevant constants,
for single particle spectra:  
\be
\frac{d^2N}{dp_t^2 d\phi} 
\sim 
\int_0^R rdr \int_0^{2\pi} d\phi_b m_t 
K_1(\beta_t) e^{\alpha_t \cos(\phi_b-\phi)},
\ee
where $\rho_t$ is the transverse flow rapidity, 
$\phi_b$ is the boost direction,
$\alpha_t=(p_t/T)\sinh(\rho_t)$, and $\beta_t=(m_t/T)\cosh(\rho_t)$.
It also assumes a uniform matter density within a cylinder, $r<R$, and a
power law transverse rapidity flow profile $\rho_t = \rho_{t,max} (r/R)^n$.
 
In such a model, the mean transverse momentum is given by expression:
\be
\la p_t \ra =
\frac{\int d\bp_t \int d\rho_t d\phi_b \rho_t^{2/n-1} p_t m_t K_1(\beta_t) 
e^{\alpha_t \cos(\phi_b-\phi)}}
{\int d\bp_t \int d\rho_t d\phi_b \rho_t^{2/n-1} m_t K_1(\beta_t) 
e^{\alpha_t \cos(\phi_b-\phi)}}.
\ee
The contribution of the transverse expansion to the two particle mean
transverse momentum correlations (both particles are from the same
$NN$ collision, the dilution factor, Eq.~3, to be applied
afterward) can be written as: 
\be
\la \dpti \dptj \ra =
\frac{
\int d\rho_t d\phi_b   \, \rho_t^{2/n-1} 
\int d\bp_{t,1}     \int d\bp_{t,2} 
( \dpti \dptj ) 
m_{t,i} K_1(\beta_{t,1}) e^{\alpha_{t,1} \cos(\phi_b-\phi_1)}
m_{t,2} K_1(\beta_{t,2}) e^{\alpha_{t,2} \cos(\phi_b-\phi_2)}
}
{
\int d\rho_t d\phi_b   \,  \rho_t^{2/n-1} 
\int d\bp_{t,1}     \int d\bp_{t,2} 
m_{t,1} K_1(\beta_{t,1}) e^{\alpha_{t,1} \cos(\phi_b-\phi_1)}
m_{t,2} K_1(\beta_{t,2}) e^{\alpha_{t,2} \cos(\phi_b-\phi_2)}
}.
\ee
This equation additionally assumes that during the expansion time (before the
freeze-out) the particles
originated from the same $NN$ collision do not diffuse far one from
another compared to the system size.
We discuss this assumption in more detail later in this paper.  
\end{widetext}

The results of the numerical calculations based on the above equations
are presented in Figs.~1 and 2 as function of $\la \rho_t^2 \ra
= \la \rho_t\ra^2 (4n+4)/(2+n)^2$. 
The results are shown for two different
velocity (transverse rapidity) profiles, $n=2$,
and $n=0.5$.  
\begin{figure}
  \includegraphics[width=0.47\textwidth]{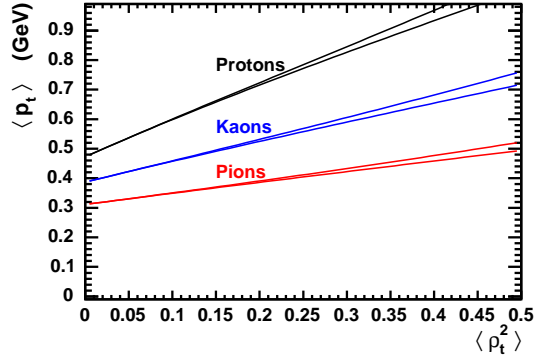}
  \caption{(color online) Mean transverse momentum 
in the blast wave calculations. $T=110$~MeV.}
  \label{fig1}
\end{figure}
\begin{figure}
  \includegraphics[width=0.47\textwidth]{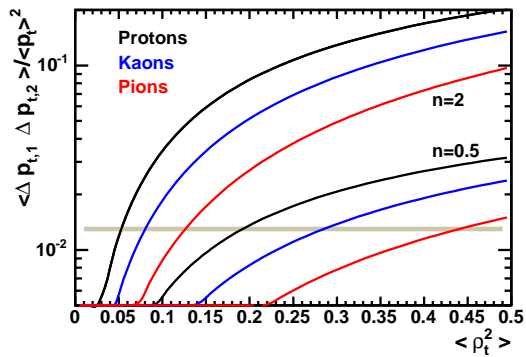}
  \caption{(color online) Two particle transverse momentum
  correlations in the blast wave calculations. $T=110$~MeV.
The gray line indicate the level of primordial correlation in
  NN-collision~\cite{isrpp} }
  \label{fig1a}
\end{figure}
One observes that indeed for all the particle types presented, $\mpt$
depends very
weakly on the actual profile. On opposite, the correlations
are drastically different for two cases presented.

{\em Rapidity correlations. Charge balance functions.}
The transverse expansion 'push' consists of many individual collisions.
It leads not only to the increase of the transverse momentum but also 
to the particle diffusion in the rapidity space. 
We do not consider 
the effect of such diffusion in this discussion concentrating only on
the effect of the transverse radial push. 
We define the balance function as
\be
B_{ab}(x_b;x_a)=\rho_{2,ab}(x_a,x_b)/\rho_{1,a}(x_a)-\rho_{1,b}(x_b).
\ee
Very roughly, $B_{ab}(x_b;x_a)$ has a meaning of a distribution 
of the 'associated' particles $b$ under condition 
of a 'trigger' particle $a$ to be found at the location $x_a$.
(The {\em charge balance function} introduced in~\cite{balance} 
can be written as
$B(b|a)=(1/2)(B_{+-}+B_{-+}-B_{++}-B_{--})$.)
Note that due to the 'normalization' to the number of 'trigger'
particles, the balance function is the same for any
superposition of independent $NN$ collisions. 
The transverse expansion leads to the narrowing of the charge balance 
function~\cite{balance2}. In our simple picture the width of the
balance function would be roughly inversely proportional to the
transverse mass as follows from the relation
\be
\Delta p_z=m_t \sinh(\Delta y) \approx m_t \Delta y.
\ee
The width decreases as the mean transverse mass increases due to 
transverse expansion. 
This effect is consistent with the experimentally observed
narrowing of the balance function with centrality~\cite{star-balance1}. 
Note that because the charge balance function is normalized to unity,
 the narrowing of the correlations means 
an increase in the
magnitude of the two particle correlation function.
In its turn it means 
an enhancement of the net charge multiplicity fluctuations 
if measured in a rapidity region compared or smaller than 
the correlation length (1--2 units of rapidity).
This observation might be an explanation for the
centrality dependence of the net charge fluctuations measured at 
RHIC~\cite{star-mult-fluc}.

{\em Azimuthal correlations.}
As all particles from the same $NN$ collisions are pushed in the same
direction (radially in the transverse plane) they become correlated 
in azimuthal space. 
The correlations can become really strong for
large transverse flow as
shown in Fig.~2 (again, for particles originated from
the {\em same} spatial position in the transverse plane).  
\begin{figure}
  \includegraphics[width=0.47\textwidth]{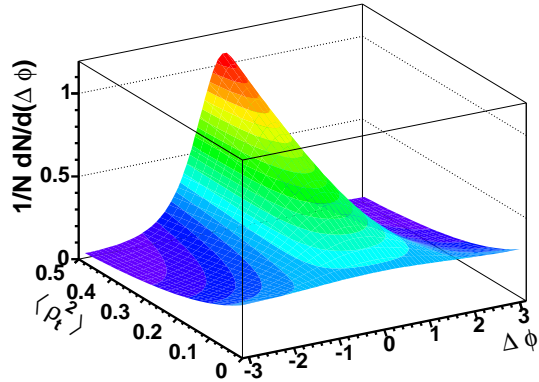}
  \caption{(color online) Two pion $\Delta \phi$ distribution as
  function of $\la \rho_t^2 \ra$ in the blast wave model. Linear velocity profile and
  $T=110$~MeV have been assumed.}
  \label{fig2}
\end{figure}
Fig.~3 shows the strength of the azimuthal correlation in terms of
the first two harmonics. 
\begin{figure}
  \includegraphics[width=0.45\textwidth]{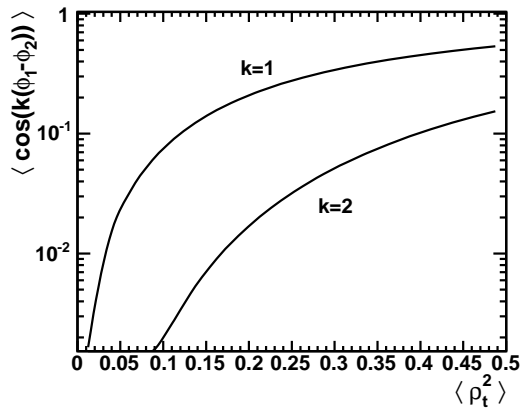}
  \caption{The average values of $\cos(\Delta\phi)$ and  $\cos(2\Delta\phi)$ for the
  distribution shown in Fig.~3.} 
  \label{fig3}
\end{figure}
$\la \cos(\Delta\phi) \ra$ reaches quite large values (even if one takes into
account a dilution factor, $D\sim 1/150$ for central Au+Au collisions). 
If the average would be
taken over all particles including the particles
from other $NN$ collisions, the momentum conservation 
contribution~\cite{momcons} can not be neglected; 
it can greatly reduce the effect. 
Note, however, that the momentum conservation effects 
are expected to be small for 
the (azimuthal) charge balance function. 
In this case both correlated particles originate from 
the same $NN$ collision and the results presented in Fig.~3 are valid.  
In our model $\cos(\Delta \phi)$ should strongly depend on 
the particle mass and its transverse momentum.  
Identified particle correlation study would be of great interest 
in this respect. 

The second harmonic in the azimuthal correlations generated by radial 
expansion is of a particular interest as it would contribute 
to the measurements of elliptic flow. 
The numbers from Fig.~3 corresponding to $\la \rho_t^2 \ra \sim0.3$ 
are comparable with the estimates 
of the strength of non-flow type azimuthal correlation estimates made 
in~\cite{star-flow-PRC}. 
Thus the azimuthal correlations generated by transverse expansion could 
be a major contributor to the non-flow azimuthal correlations.
More importantly, this contribution would depend on centrality (following
the development of radial flow), unlike many other non-flow effects.
Note, however, that the typical elliptic flow centrality 
dependence (rise and fall) is different
from that of the correlations due to transverse 
radial expansion (continuous 
increase of $N_{coll} \cdot \la \cos(2 \Delta \phi) \ra$).  

{\em Thermalization time. Diffusion.}
The magnitude of the correlations due to transverse expansion should 
be sensitive to the system thermalization time and the particle
diffusion in the thermalized matter during the expansion; 
the azimuthal correlations would be the most interesting/useful
 for such a study.
The dependence on the thermalization time comes from the following. 
Somewhat exaggerating, one can imagine that just after the collision, 
newly produced particles ('freed' partons) experience free 
streaming for some period of time before the thermalization happens. 
After that moment the particles become frozen into hydrodynamic type 
cell and an expansion starts with all the consequences 
discussed in the first part of this paper. 
The effect of the free streaming phase would be 
a diffusion in the transverse plane 
of the particles created initially at the same position 
(of the typical hadronic size). 
Such a diffusion would lead to a broadening
of the azimuthal correlations. 
Many current estimates give the thermalization time of the order of
or smaller than 1~fm/c~\cite{hydroreview}. 
Such a short thermalization time would be difficult to observe, 
but in any case the corresponding measurements 
would be of great interest giving an independent limit
to this important parameter.

{\em Single jet tomography?}
We could go even further with our speculations. Jet tomography
of nuclear collisions is a popular subject based on 
the jet quenching phenomena. 
For such study it would be very useful to find
an observable that is correlated to the space point where 
the hard collision occurred. 
It seems that the correlations due to
transverse expansion provide such a possibility.
One has to correlate the jet (high $p_t$ hadron) yield
with mean transverse momentum of particles taken at different rapidity
(but better at similar azimuthal angle). 
In the same $NN$ collision where the high $p_t$ particle is emitted, the soft
particles are produced as well. Those
soft particles experience the transverse 'push' corresponding
to the spatial position in the transverse plane where 
the original $NN$ collision happens to be. 
Then the mean transverse momentum of the associated
particles would provide the information on how close to the center 
of the system the collision occurred.      

{\em Summary.} Expansion of the dense system
created at the initial stage of a high energy nucleus-nucleus collision
leads to strong position-momentum correlations.  
As all particles from the same $NN$ collision are produced at the same
position in the transverse plane, they get
a similar radial push during the expansion stage.
It creates rapidity, azimuthal angle, and
transverse momentum correlations. 
The correlations extend over wide rapidity range. 
The picture is not boost-invariant, 
as the initial geometry of the source and particle densities 
change with rapidity leading to a rapidity dependent radial expansion.

The above described picture of $AA$ collisions has many
interesting observable effects, only a few mentioned in this paper. 
The picture become even richer if one looks at the 
identified particle correlations.
Many  questions require a
detailed model study, but the approach opens a potentially very interesting
possibility to address the initial conditions and the subsequent
evolution of the system created in an $AA$ collision.  

\acknowledgments{
Fruitful discussions with R.~Bellwied, S.~Gavin, and C.~Pruneau, and
critical comments of U.~Heinz are gratefully acknowledged. 
This work was supported in part by the
U.S. Department of Energy Grant No. DE-FG02-92ER40713.
}

\bibliographystyle{unsrt}
 \end{document}